\documentclass[manuscript]{acmart}

\usepackage{booktabs}
\usepackage{microtype}
\usepackage{graphicx}
\usepackage{cleveref}
\graphicspath{{../figures/}}

\acmConference[CSCW '26]{ACM Conference on Computer-Supported Cooperative Work and Social Computing}{October 10--14, 2026}{Salt Lake City, UT, USA}
\acmYear{2026}
\acmISBN{}
\acmDOI{}

\author{Shiyao Wei}
\affiliation{%
  \institution{Florida State University}
  \city{Tallahassee}
  \country{United States}
}
\email{sw22b@fsu.edu}

\author{Ran Bi}
\affiliation{%
  \institution{SAS Institute}
  \city{Cary}
  \country{United States}
}
\email{ran.bi@sas.com}

\title{Does Discussion Matter? Interaction Dynamics in Transparent Peer Review}

\begin{document}

\begin{abstract}
Open peer review is advocated as a way to make review more dialogic, yet evidence on whether it changes participant behavior remains scarce. We compare two transparent review designs: Nature Human Behaviour, an editor-mediated model, and International Conference on Learning Representations 2023, a discussion-based model. Treating cross-mentions, moments when one participant explicitly names another beyond the immediate dyad, as the analytic unit, we combine social network analysis of 56 NHB articles and 371 ICLR notable papers with qualitative coding grounded in Knowledge Building theory. Structurally, NHB forms hub-and-spoke networks with zero reviewer-to-reviewer reciprocity, whereas ICLR forms partial-mesh networks with significantly more direct cross-referencing. Functionally, the distribution of communicative functions is similar across venues, but who performs them differs: editors aggregate at NHB, while authors and reviewers absorb that labor at ICLR. The designs differ most in how communicative labor is distributed across roles, suggesting platforms should support it as explicit, built-in functions.

\end{abstract}




\maketitle

%% ----------------------------------------------------------------
\section{Introduction}
%% ----------------------------------------------------------------

Open peer review has been advocated, in part, as a way to improve interaction among participants in the review process. Open interaction, the facilitation of direct, reciprocal exchange among authors, editors, and reviewers, is positioned as a mechanism for moving peer review beyond isolated, one-directional evaluation toward a multi-party conversation in which knowledge is collectively negotiated \cite{ross2017open}.

Yet we have little empirical evidence about whether this actually happens.
Research has focused on the exchange between a single reviewer and the authors, examining review quality, tone, and specificity, or the relationship between reviewer scores and editorial decisions~\cite{grimaldo2018fragments,kennard2022disapere,Liu_Yuan_2025,shi2024we,Smith_Kennard_Du_McFarland_2025}. What happens across these pairs, the moments when one participant takes up what another said to someone else, has received less attention~\cite{tennant2020limitations}.

This research gap mirrors a design gap in publication infrastructure. We use \emph{transparent} to describe any venue that publishes its review documents, and \emph{discussion-based} for the narrower subset that additionally permits direct multi-party reply. Nature Human Behaviour (NHB) exemplifies a mediated transparent model, in which the editor synthesizes reviewer feedback into a decision letter that authors answer in a single rebuttal; the International Conference on Learning Representations (ICLR) runs discussion-based review on the OpenReview platform, where reviewers, authors, and program chairs respond to each other directly in an open thread. Yet evidence on what these designs actually produce remains limited, and Tennant and colleagues~\cite{tennant2017multi} have called for empirical research on whether opening review changes reviewer behaviors.

This study takes up that call. Drawing on Knowledge Building theory~\cite{scardamalia1994computer}, we analyze the discourse functions that \emph{cross-mentions} (moments in which one participant explicitly references another beyond the immediate dyad) serve in these two open review systems. We ask two research questions: (1)~what cross-mention dynamics are present in transparent peer review at NHB and ICLR 2023, and (2)~what do cross-mention dynamics imply for collaborative knowledge creation?

%% ----------------------------------------------------------------
\section{Related Work}
%% ----------------------------------------------------------------

Most empirical work on peer review treats it as an evaluation mechanism, reporting on reviewer agreement, bias, and editorial outcomes~\cite{grimaldo2018fragments,Smith_2006,Smith_Kennard_Du_McFarland_2025}. Work that treats review as situated language use annotates discourse moves within reviewer--author threads~\cite{kennard2022disapere,Liu_Yuan_2025}, and quantitative studies of OpenReview venues measure discussion length, score change, and participation at scale~\cite{kargaran2025insights,miyao2019does,tran2020open}. The analytic unit throughout is the single exchange or the paper-level aggregate. A parallel computational literature detects references within scholarly text (entities and concepts~\cite{luan-etal-2018-multi,cattan2021scico}, citation function~\cite{teufel2006annotation,jurgens-etal-2018-measuring}, argumentative structure in peer review~\cite{hua2019argument,kang2018dataset,cheng2020ape}), but none of it tracks reference to the participants themselves. Cross-mention analysis shifts the analytic gaze from within-exchange discourse to between-exchange connections.

Our coding scheme builds on Stahl's account of group cognition~\cite{stahl2006group}, which locates meaning-making in the interaction between participants rather than in individual minds. Stahl separates coordinative work (turn-taking, referencing, topic management) from epistemic moves (proposing, critiquing, synthesizing), and Scardamalia and Bereiter's Knowledge Building theory~\cite{scardamalia1994computer,bereiter2014knowledge,van2009distinguishing} distinguishes knowledge sharing, individuals transmitting what they already know, from knowledge building, the collaborative improvement of ideas as shared objects. Both dynamics may occur on either platform. What we ask is how often each occurs and who drives it. Whether cross-mentions serve coordinative or epistemic functions, and whether they index individual sharing or collaborative building, is the empirical question our coding scheme answers.
%% ----------------------------------------------------------------
\section{Methods}
%% ----------------------------------------------------------------
\subsection{Data and Cross-Mention Extraction}
From NHB, whose transparent review process publishes the full editor--reviewer--author correspondence alongside each accepted article, we collected all 56 articles with publicly available 2023 peer review records (169 cross-mention events). From ICLR 2023 we retrieved all 371 notable papers (top-5\% Oral tier, $n{=}91$; top-25\% Spotlight tier, $n{=}280$) via the OpenReview API (148 events). We restricted ICLR to the top tier because NHB publishes records only for accepted manuscripts. No personally identifiable information beyond publicly attributed reviewer roles is used.

We identified cross-mentions in two rounds. First, one author read every paper's review and annotated every instance in which a writer named a third participant, recording sender and recipient. Second, we ran a keyword search on the token \emph{reviewer} across every reply text field. Informal short forms and plural collective references were excluded, leaving only mentions resolvable to a specific reviewer. The extraction patterns and exclusion audits are detailed in Appendix~\ref{app:extraction}.

\subsection{Social Network Analysis}
We built a directed graph per paper. Nodes are the actors observed in that paper's record (author, reviewers, Editor or Program Chair, public commenter), and each edge $A \to B$ encodes a cross-mention, weighted by mention frequency. No structural edges are added, since cross-mentions are the only unit comparable across venues. We computed three per-paper metrics: density ($\delta$), the share of possible reviewer-to-reviewer connections that actually occur; reciprocity, the fraction of reviewer-to-reviewer references that are returned; and in-degree centrality, the number of participants referencing a given reviewer, which we read as marking emergent opinion leaders. Because extraction matches named-reviewer references only, edges always terminate at reviewer nodes, and non-reviewer roles appear exclusively as senders; in-degree is therefore computed over reviewer nodes only. All analyses used Python and \texttt{NetworkX}.

\subsection{Qualitative Coding}
To understand the collaborative knowledge creation behind the cross-mentions, we open-coded the 169 NHB instances. One author developed the codebook from the data; two coders then double-coded 18\% (30 instances, Cohen's $\kappa{=}0.904$) before coding the remainder independently. The same procedure was applied to the 148 ICLR 2023 instances (double-coded subset $\kappa{=}0.734$); remaining disagreements were resolved through discussion.

The codebook groups five functions into two families. \emph{Coordinative} moves manage multi-party logistics: \emph{aggregation} (synthesizing multiple reviewers' feedback into one message) and \emph{redirection} (pointing a reviewer to an existing response or another reviewer's comment). \emph{Epistemic} moves engage with knowledge claims: \emph{alignment} (constructing or exhibiting consensus), \emph{attribution} (crediting specific reviewers), and \emph{conflict} (articulating disagreement or recruiting others' concerns to challenge a revision). Coders assigned the single most applicable code per instance. To compare the marginal code distribution across venues, we report a paper-level permutation $\chi^{2}$ test (5{,}000 permutations of venue labels across papers) alongside the conventional i.i.d.\ statistic.

\subsection{Network Comparison Test}
To test whether role-pair edge frequencies differ across venues, we applied a paper-level permutation Network Comparison Test (NCT)~\cite{vanBorkulo2022} over five role-pair edge types (\Cref{tab:nct}). The NCT is computed on the subset of papers containing $\geq 1$ cross-mention (49 NHB, 108 ICLR) and therefore tests how activity is distributed across role-pairs given that any occurred; population-level prevalence is reported at the start of \S\ref{sec:findings}. Permutation mechanics, multiple-comparison correction, and global statistics are detailed in Appendix~\ref{app:nct}.

%% ----------------------------------------------------------------
\section{Findings}
\label{sec:findings}
%% ----------------------------------------------------------------

NHB articles are far more likely to contain at least one cross-mention event (87.5\%) than ICLR oral (33.0\%) or spotlight (27.9\%) papers, and they average 1.15 cross-mentions per reviewer against 0.14 and 0.10 respectively (counting mentions from any sender), reflecting the active editorial synthesis role in mediated review. Mean per-paper density on the reviewer$\to$reviewer subgraph is nonetheless low in all three corpora (NHB 0.057, ICLR oral 0.033, spotlight 0.018). The structure of those cross-mentions differs fundamentally. NHB's activity is dominated by unidirectional references, while ICLR produces sparser but individually targeted reviewer-to-reviewer mentions.

\subsection{RQ1: Cross-Mention Dynamics}
In NHB, cross-mention activity is dominated by editor synthesis messages and author rebuttals that reference named reviewers, so the Editor and the Author account for the bulk of inward flow toward reviewer nodes.
Direct reviewer-to-reviewer mentions exist but are rare (15 directed edges across 49 articles, concentrated in 13 papers), and every one of them is unidirectional.

\begin{figure}[h]
  \centering
  \includegraphics[width=0.7\linewidth,trim={0 48pt 0 30pt},clip]{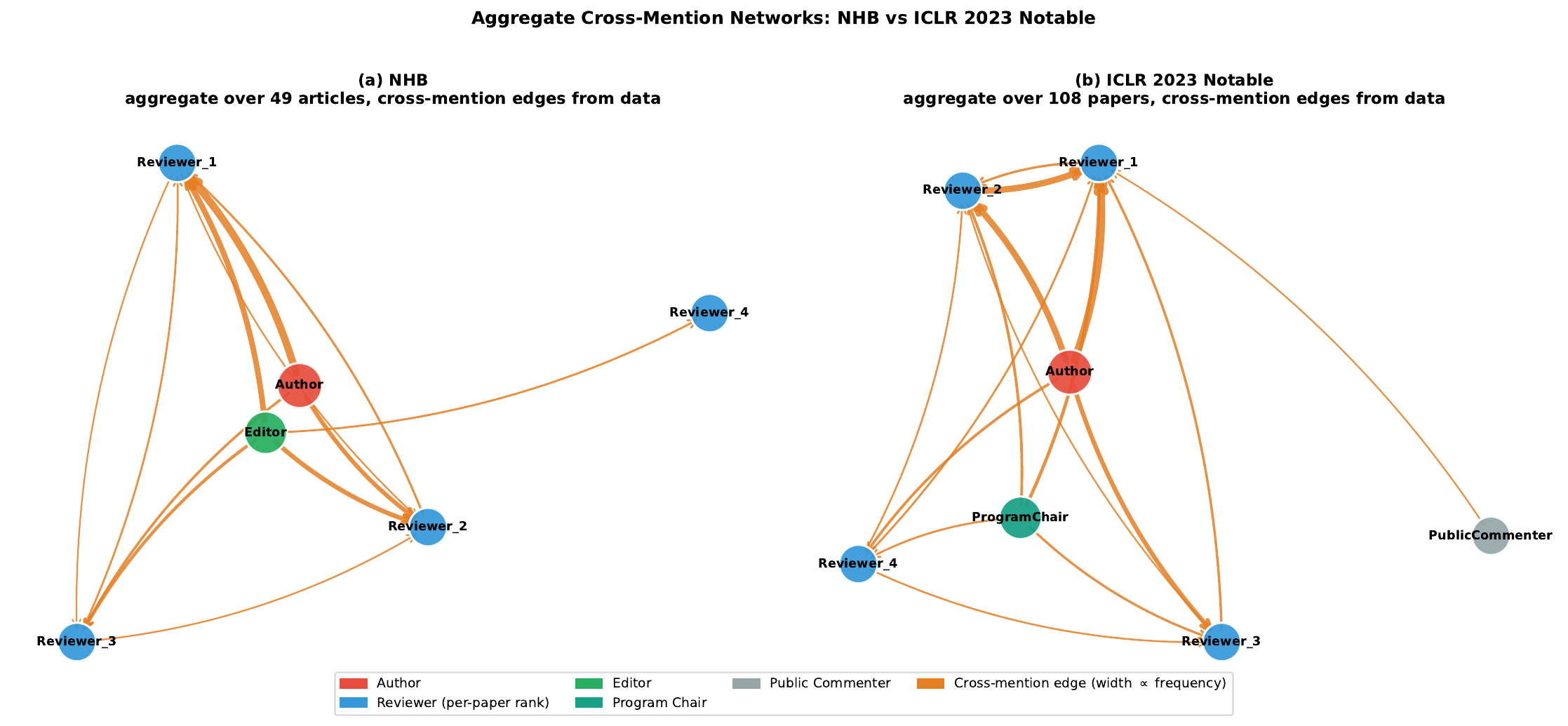}
  \Description{Two side-by-side aggregate directed network graphs. All arrows terminate at reviewer nodes by extraction design. Left (NHB): Editor and Author are the dominant senders pointing inward to reviewer nodes; a few reviewer-to-reviewer arrows appear but they are pooled across papers and no within-paper reciprocal pair exists. Right (ICLR): Author is still the most active sender targeting individual reviewers, with substantive reviewer-to-reviewer edges including reciprocal pairs, plus Program Chair messages mentioning individual reviewers.}
  \caption{Aggregate cross-mention networks pooled across papers with $\geq 1$ named-reviewer event: (a)~NHB, (b)~ICLR 2023 Notable. Edge width $\propto$ pooled mention frequency.}
  \label{fig:aggregate}
\end{figure}

In ICLR, across the 108 papers with $\geq 1$ cross-mention, we observe 75 direct reviewer-to-reviewer edges, 8 of which are reciprocal (in 4 papers); 0 such reciprocal pairs occur in NHB (mean reviewer-to-reviewer reciprocity: 1.8\% oral, 0.7\% spotlight, 0\% NHB).
\Cref{fig:aggregate} contrasts the two aggregate networks. Repeated cross-mentions converge on a hub-and-spoke topology in NHB, where editors and authors point inward to a few focal reviewers with no reviewer-to-reviewer return path, versus a mildly reciprocal partial mesh in ICLR, where a small number of exchanges close into reciprocal pairs.

\begin{figure}[h]
  \centering
  \includegraphics[width=0.7\linewidth,trim={0 0 0 24pt},clip]{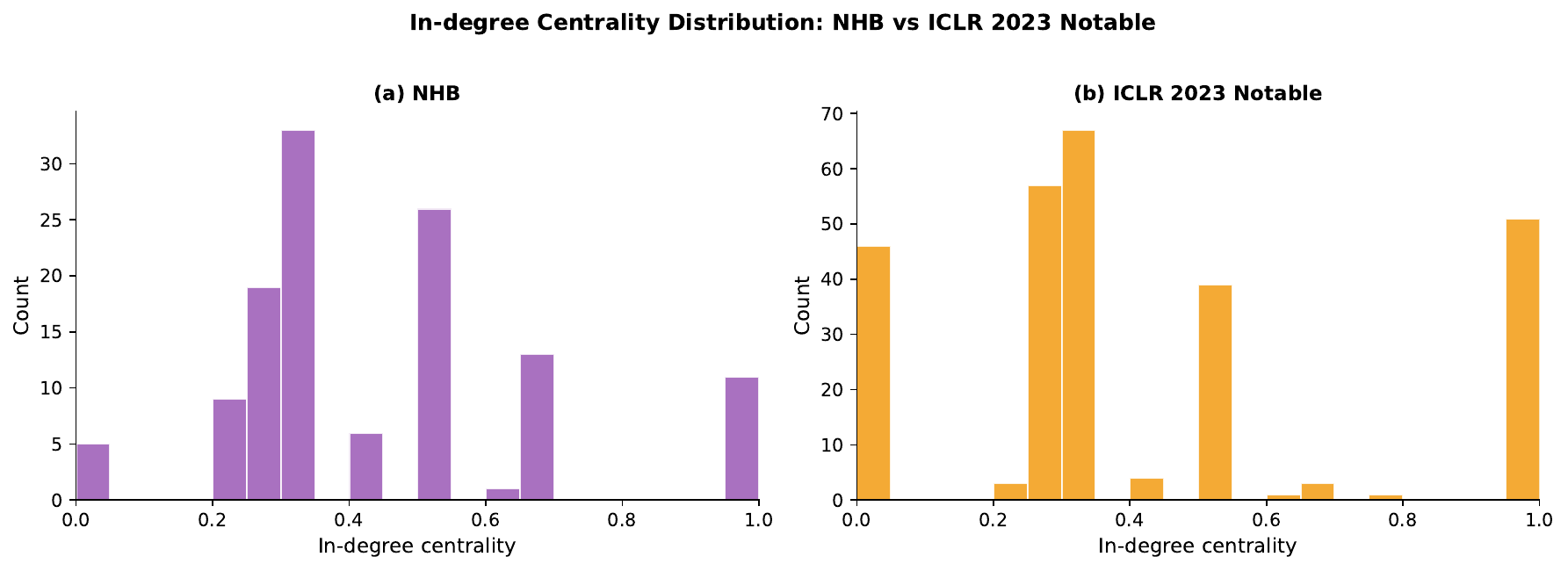}
  \Description{Two-panel histogram of reviewer-node in-degree centrality, restricted to reviewer nodes because only reviewers receive cross-mentions in this corpus. Panel (a) NHB and panel (b) ICLR 2023 Notable both show strongly right-skewed distributions, with the bulk of reviewers near zero and a thin tail extending toward 1.}
  \caption{In-degree centrality distribution for reviewer nodes in the per-paper cross-mention subgraphs: (a)~NHB, (b)~ICLR 2023 Notable.}
  \label{fig:centrality}
\end{figure}

\Cref{fig:centrality} shows in-degree centrality for reviewer nodes.
Both distributions are right-skewed: most reviewers receive few mentions, while a few attract many, functioning as potential opinion leaders.
The skew is more pronounced in ICLR, where co-reviewers also converge on the same focal reviewer in addition to authors and chairs, whereas in NHB the convergence comes mainly from the Editor and Author.

The NCT confirms that the two networks are globally non-equivalent ($M{=}2.12$, $p{<}.001$; $S{=}3.80$, $p{<}.001$) and decomposes that difference into two findings (\Cref{tab:nct}, Appendix~\ref{app:nct}).

First, mediating roles flip across venues. The editor~$\to$~reviewer edge, the largest difference observed, is the statistical signature of NHB's hub-and-spoke topology, and ICLR Program Chairs fulfill a similar but weaker coordination function. NHB authors reference reviewers at roughly 1.7$\times$ the ICLR rate, because a gated response letter requires naming each reviewer explicitly whereas a threaded discussion makes the referent visually co-present.

Second, reviewer-to-reviewer cross-referencing is significantly higher in ICLR. Reviewers cross-reference one another at roughly twice the per-paper rate of NHB (0.69 vs.\ 0.33 mentions per paper, 95\% bootstrap CI of $|\Delta|$ $[0.14, 0.59]$, Cliff's $\delta{=}0.26$). Because the venues differ in both discussion affordance and disciplinary review culture (\S\ref{sec:discussion}), we read this as consistent with channel design shaping cross-referencing behavior rather than as causal evidence.

\subsection{RQ2: Cross-Mention Dynamics and Collaborative Knowledge Creation}
Qualitative coding of 49 NHB articles revealed five functional categories whose distribution differs markedly by participant role (\Cref{tab:functions}).

\begin{table}[h]
\caption{Distribution of cross-mention functions by participant role and venue (NHB $N{=}169$, plural references excluded; ICLR 2023 $N{=}148$, one public-commenter mention omitted for brevity).}
\label{tab:functions}
\resizebox{\columnwidth}{!}{%
\begin{tabular}{lrrrrrr}
\toprule
& \multicolumn{3}{c}{NHB} & \multicolumn{3}{c}{ICLR 2023} \\
\cmidrule(lr){2-4}\cmidrule(lr){5-7}
Function & Editor ($n{=}54$) & Author ($n{=}101$) & Reviewer ($n{=}14$) & Prog.\ Chair ($n{=}20$) & Author ($n{=}59$) & Reviewer ($n{=}68$) \\
\midrule
Aggregation  & 52 (96\%) & 3 (3\%)   & 0 (0\%)   & 10 (50\%) & 42 (71\%) & 5 (7\%)   \\
Attribution  & 1 (2\%)   & 27 (27\%) & 5 (36\%)  & 5 (25\%)  & 9 (15\%)  & 18 (26\%) \\
Redirection  & 1 (2\%)   & 42 (42\%) & 1 (7\%)   & 2 (10\%)  & 7 (12\%)  & 27 (40\%) \\
Alignment    & 0 (0\%)   & 28 (28\%) & 6 (43\%)  & 2 (10\%)  & 1 (2\%)   & 16 (24\%) \\
Conflict     & 0 (0\%)   & 1 (1\%)   & 2 (14\%)  & 1 (5\%)   & 0 (0\%)   & 2 (3\%)   \\
\bottomrule
\end{tabular}}
\end{table}

Editors cross-mentioned almost exclusively to aggregate (96\%), synthesizing feedback from multiple reviewers into unified decision letters: ``Our reviewers request several new analyses. Reviewer 1 asks that you directly compare \ldots{} Reviewer 2 asks that you conduct additional empirical analyses\ldots''

Authors deployed cross-mentions across a broader range. Redirection (42\%) managed redundancy by pointing reviewers to responses already provided elsewhere (``\ldots{}in response to this and Reviewer 2's similar comment (Sec 3.3)''). Alignment (28\%) constructed reviewer consensus to position a single response as satisfying the collective, and Attribution (27\%) credited specific reviewers for contributions that shaped the revision.

Reviewers cross-mentioned primarily to align with another reviewer's stance (43\%) or to attribute a point to a specific colleague (36\%), revealing a division of epistemic labor within the panel: ``One more thing that I'd suggest authors should include (after reading Reviewer \#2's comment)\ldots''
Conflict was rare overall (2\%) but concentrated among reviewers (14\% of their cross-mentions).

The same codebook was applied to the 148 ICLR 2023 instances. The marginal distribution of the five functions shows similar trends across venues, and we do not observe a significant difference under the paper-level permutation $\chi^{2}$ test ($\chi^{2}(4){=}3.10$; i.i.d.\ $p{=}0.54$; cluster-robust $p{=}0.71$; Cram\'er's $V{=}0.10$); absence of evidence is not evidence of equivalence, and a confirmatory equivalence test is left for follow-up. What does differ is the role that performs each function (\Cref{tab:functions}).

Two role differences stand out. Aggregation flips from editor to author. In NHB, 95\% of aggregation is editor-produced (52/55); in ICLR, with no editorial presence, authors take over the synthesis function (71\% of their cross-mentions, 42/59): ``The main limitation raised by Reviewer xxxx and xxxx was that the ability to use Imagenet-X\ldots'' Alignment becomes a reviewer-driven move in ICLR: NHB reviewer-driven alignment occurs but at low absolute frequency (6 events, $43\%$ of reviewer cross-mentions), while ICLR reviewers themselves drive 80\% of alignment events (16/20), publicly echoing each other's stances. Conflict remains rare in both venues and reviewer-concentrated.

%% ----------------------------------------------------------------
\section{Discussion}
\label{sec:discussion}
%% ----------------------------------------------------------------
Across two venues, the marginal distribution of the five discourse functions shows similar trends, while the agent of each move tracks platform affordances. Together with the structural divergence between hub-and-spoke and partial mesh, this pattern is consistent with discussion-based review relocating who performs multi-party discourse work rather than producing different kinds of discourse, offering preliminary evidence for the call by Tennant et al.\ \cite{tennant2017multi}.

Because the labor of synthesis is reassigned to authors, the design question shifts from whether to open peer review to how to support the communicative work openness exposes. Aggregation needs a designated home. NHB locates it in the editor, while ICLR leaves it to authors, a burden likely borne disproportionately by less-resourced teams \cite{nigatu2026unweirding}. ICLR reviewers already perform alignment unprompted, yet direct reviewer-to-reviewer negotiation stays rare even where permitted \cite{tennant2020limitations}. These observations converge on a pre-rebuttal consensus digest: a structured form the reviewer panel completes jointly before authors respond, recording agreements, disagreements, and each reviewer's most decisive concern.

Two limitations bound these claims. First, cross-mention captures only the explicit end of multi-party coordination: NHB's letter-based format surfaces coordination as named reference, whereas OpenReview's threaded interface lets a reviewer respond to another in place without naming them. We retain the mention-only rule as the only edge definition symmetric across the venues, so ICLR reviewer$\to$reviewer counts and reciprocity are a lower bound (unmeasured reply edges would only strengthen the NCT finding). Second, the corpora limit generalizability: to match NHB's accept-only publication policy we sampled only ICLR's notable tiers, leaving rejected and borderline submissions out of scope; and because NHB (behavioral science) and ICLR (machine learning) differ in both discussion affordance and disciplinary review culture, we cannot isolate platform design from venue-specific norms. Replication across venues with contrasting discussion policies is future work.

%% ----------------------------------------------------------------
\section{Conclusion}
%% ----------------------------------------------------------------
Treating cross-mentions as the analytic unit makes visible the multi-party work that dyadic accounts of peer review obscure. NHB's mediated model yields hub-and-spoke networks with zero reviewer-to-reviewer reciprocity; ICLR's discussion-based model yields partial-mesh networks with significantly more direct cross-referencing. Yet the five discourse functions show similar trends across venues; what differs is who performs them. Subject to the disciplinary--platform confound, and measuring no review-quality outcomes, these descriptive results suggest the visible difference between designs lies in how communicative labor is distributed, not in its content~\cite{tennant2017multi}. Platforms should treat that labor as first-class functions to be scaffolded, rather than leaving it to whichever participant is left holding it.
%% ----------------------------------------------------------------
\begin{acks}
We thank OpenReview and Nature Human Behaviour for making peer review data publicly accessible.
The authors used Claude 4.6 for language editing only; it generated no ideas, concepts, or references, and all authors carefully reviewed the manuscript and take full responsibility for its final content.
\end{acks}

\bibliographystyle{ACM-Reference-Format}
\bibliography{references}

\appendix
\section{Cross-Mention Extraction Details}
\label{app:extraction}
The keyword search was anchored on the token \emph{reviewer} across every reply text field (reviews, rebuttals, program chair messages, public comments), using venue-specific patterns: ordinal forms (\emph{Reviewer 2}) for NHB, resolved against the per-paper roster, and unique handles (\texttt{Reviewer\_xxxx}) for ICLR. Two authors assigned NHB sender and recipient roles from each comment's position in the published review history. Informal short forms (\texttt{R1}, ``the other reviewer'') were excluded: an audit found them in only 1.9\% of ICLR replies, almost all false positives (e.g., \texttt{Table R2}). Plural collective references were excluded from both corpora for symmetry, leaving only mentions resolvable to a specific reviewer.

\section{Network Comparison Test Mechanics}
\label{app:nct}
The test statistic is the absolute difference in mean mentions-per-paper between venues; the null distribution comes from 5{,}000 permutations that reassign papers (not individual events) to venue groups, making the test robust to within-paper event clustering. We applied Benjamini--Hochberg FDR correction~\cite{benjamini1995controlling} across the five edge tests at $\alpha{=}0.05$ and computed two global statistics, $M$ (the maximum single-edge difference) and $S$ (the sum of edge differences), against the same permutation distribution. \Cref{tab:nct} reports the per-edge results.

\begin{table}[h]
\caption{NCT results: role-pair edge comparison between NHB ($n{=}49$) and ICLR ($n{=}108$) via 5{,}000 paper-level permutations. $\bar{x}$ = mean mentions per paper.}
\label{tab:nct}
\centering\small
\begin{tabular}{lrrrl}
\toprule
Edge type & $\bar{x}_{\text{NHB}}$ & $\bar{x}_{\text{ICLR}}$ & $|\Delta|$ & $p_{\text{FDR}}$ \\
\midrule
editor $\to$ reviewer         & 2.12 & 0.00 & 2.12 & $<.001$ \\
author $\to$ reviewer         & 2.20 & 1.30 & 0.91 & $.017$  \\
program chair $\to$ reviewer  & 0.00 & 0.39 & 0.39 & $.007$  \\
reviewer $\to$ reviewer       & 0.33 & 0.69 & 0.37 & $.011$  \\
\bottomrule
\end{tabular}
\smallskip\par\noindent{\footnotesize public commenter $\to$ reviewer omitted (ICLR $\bar{x}{=}0.009$, $p{=}1.00$).}
\end{table}

\end{document}